\documentclass{aastex6}

\def\au{{\rm au}}
\begin{document}


\title{FORMATION OF WIDE-ORBIT GAS GIANTS NEAR THE STABILITY LIMIT IN MULTI-STELLAR SYSTEMS}


\author{A. Higuchi}
\affil{Department of Earth and Planetary Sciences, Faculty of Science, 
  Tokyo Institute of Technology, Meguro, Tokyo 152-8551, Japan}
\and
\author{S. Ida}
\affil{Earth-Life Science Institute,
  Tokyo Institute of Technology, Meguro, Tokyo 152-8550, Japan}


\begin{abstract}
    We have investigated the formation of a circumstellar wide-orbit gas giant planet in 
    a multiple stellar system.
    We consider a model of orbital circularization for the core
    of a giant planet after 
    it is scattered from an inner disk region by a more massive planet, which was proposed 
    by \citet{khi14}.
    We extend their model for single star systems to binary (multiple) star systems, 
    by taking into account tidal truncation of the protoplanetary gas disk by a binary companion.
    As an example, we consider wide-orbit gas giant in a hierarchical triple system, HD131399Ab.
    The best-fit orbit of the planet is that with semimajor axis $\sim 80$ au and eccentricity $\sim 0.35$.
    As the binary separation is $\sim 350$ au, it is very close to the stability limit, which is puzzling.
    With the original core location $\sim 20$-30 au, the core (planet) mass $\sim 50 M_{\rm E}$ and the disk 
    truncation radius $\sim 150$ au, our model reproduces the best-fit orbit of HD131399Ab.
    We find that the orbit after the circularization is usually close to the stability limit against 
    the perturbations from the binary companion, because the scattered core accretes gas from the truncated disk.
    Our conclusion can also be applied to wider or more compact binary systems if the separation is not 
    too large and another planet with $\ga$ 20-30 Earth masses that scattered the core existed in inner 
    region of the system.
\end{abstract}

\section{INTRODUCTION}

Most of the stars in our Galaxy are members of binary or triple star systems.
Many exoplanets have been discovered in circumstellar orbits in multiple star systems 
by transit and radial velocity surveys.
Although the circumstellar orbits of planets can be destabilized by secular perturbations from a binary companion,
they are stable if their orbital radii are less than 
  a critical value.
  \citet{holman99} derived a fitting formula for the critical separation for stability as
  \begin{equation}
    a_{\rm c} \simeq (0.46 - 0.38\mu - 0.63 e_{\rm b} +0.59 \mu e_{\rm b})a_{\rm b},
    \label{eq:a_cr}
  \end{equation}
  where $\mu$ is the binary companion mass scaled by
  the sum of the host and companion stars, $e_{\rm b}$ and $a_{\rm b}$ are
  the eccentricity and 
  semimajor axis, respectively,
  of the binary companion orbit, and circular orbits are assumed for the planets.
  For an equal-mass binary pair ($\mu = 0.5$) with a circular binary orbit,
  $a_{\rm c} \simeq 0.27 a_{\rm b}$.

Wide-orbit extrasolar gaseous giant planets in
nearly circular orbits have been detected by direct imaging observations
for several systems \citep[e.g.,][]{k08, m08, k13}.
One of the latest announcements is a $(4 \pm 1) M_{\rm Jup}$ planet in the HD131399
triple star system \citep{w16}.
The primary star, HD131399A, is an A-type star with mass $\simeq 1.82 M_\odot$,
which the discovered planet (HD131399Ab) orbits.
The close binary of a G-type star with mass $\simeq 0.96 M_\odot$ (HD131399B) and a K-type star 
with mass $\simeq 0.6 M_\odot$ (HD131399C) is orbiting HD131399A
with semimajor axis of $\sim 350 \, \au$ and eccentricity of $\sim 0.13$ \citep{w16}.
  With these values, Eq.~(\ref{eq:a_cr}) shows $a_{\rm c} \simeq 0.24 a_{\rm b} \simeq 84$ au.
The estimated semimajor axis and eccentricity of the planet HD131399Ab are 
$82^{+23}_{-27}$ au and $0.35\pm 0.25$ \citep{w16}.
  Although Eq.~(\ref{eq:a_cr}) is for planets in circular orbits, the estimated planetary semimajor axis
  is close to $a_{\rm c}$ and
the planet's orbit would be marginally stable or possibly unstable on long timescales \citep{w16,veras17}. 
Although the planet could be a background star (Nielsen E. L. et al., submitted; arXiv:1705.06851), if it is a planet,
how to form such a marginally stable world in a multiple planetary system is a challenge for planet formation theory.

Because in situ formation of gas giant planets is difficult at $\sim 80\,\au$ with the conventional
core accretion scenario \citep[e.g.,][]{il04}, \citet{w16} discussed the following three possibilities:
(i) planet$-$planet scattering in the star A system, (ii) formation of a circumbinary planet around 
the B/C pair followed by transfer to the A system, and 
(iii) formation of the planet 
around 
either of the stars and evolution in the course of stellar orbital evolution of the triple stellar system
before settling down to the current stellar configuration.
For model (i), an additional unseen massive planet is required in the inner region.
Because the eccentricity after the scattering must be close to unity, 
an eccentricity damping mechanism for the scattered planet is also required. 
Although models (ii) and (iii) are not ruled out, it is not clear what evolutionary path 
leads to the current planetary and stellar orbital configurations.

Here, we are focused on the scattering model and discuss how
a gas planet with baffling features (a wide-separation 
moderate-eccentric orbit close to the stability limit by perturbations from a companion(s))
such as HD131399Ab can be reproduced.
Planet scattering by a more massive planet 
can significantly increase the apoastron distance,
but not the periastron distance.
To explain a wide-orbit gas giant with moderate orbital eccentricity, a mechanism 
to lift the periastron distance, equivalently, to damp the eccentricity without significant semimajor axis damping is required.

The Lidov$-$Kozai mechanism \citep{k62, l62} is one of the possible eccentricity variation mechanisms.
The secular perturbation from the binary companions orbiting around HD131399A 
can induce large amplitude oscillations of the eccentricity $e$ and inclination $i$
of HD131399Ab without changing the semimajor axis due to conservation of
the vertical component of the orbital angular momentum ($\propto \sqrt{1-e^2}\cos i$) around A,
if the initial argument of periastron is proper.
In this case, however, the proper range of the initial argument of periastron is narrow.
Furthermore,  even if the narrow range of initial conditions is satisfied,
the periastron distance repeatedly becomes smaller according to the oscillation,
and HD131399Ab's orbit can also be repeatedly perturbed by the massive planet in the inner region that
scattered HD131399Ab.
It is not clear if HD131399Ab's orbit can be stable during the age of the host star. 

 Another possibility is dynamical friction from the protoplanetary disk gas
 if it exists after the scattering.
 \citet{bk14} investigated the orbital evolution of eccentric planets by dynamical friction. 
They found that the periastron distance is increased by a factor of at most four
 times in the case of uniform disk gas depletion.
In the case of inside-out depletion, dynamical friction from the outer part of the disk
contributes to significant lift-up of the periastron distance,
but it is unlikely that the inside-out disk gas depletion continues up to the $\sim 50\,\au$
that is required to reproduce the orbit of HD131399Ab.
 
The model proposed by \citet{khi14} is also based on planet$-$planet scattering,
but considers scattering of a solid core and the eccentricity damping during gas accretion onto the core
to form a gas giant.
The gas accretion is most efficient when the core is near the apoastron
in the highly eccentric orbit after the scattering.
Even if the disk gas surface density of $\propto r^{-3/2}-r^{-1}$, 
the periastron distance is significantly increased.
This model does not require a gas giant in an inner region more massive than the wide-orbit gas giant in the final state.
This model is a promising option to reproduce the current orbital configuration of HD131399Ab.

The model also naturally explains why the orbit is close to the stability limit.
\citet{khi14} considered the formation of a wide-orbit gas giant around a single star
and found that the final orbit is regulated by the radius of the disk's outer edge,
because gas accretion onto the planet does not occur beyond the disk's outer edge.
  When a stellar companion exists, the circumstellar disk is truncated 
  at $\sim 1/3$ of the binary separation 
  by the perturbations from the companion with $e_{\rm b} \sim 0.1$,
  and the truncation radius becomes larger for
  a smaller-mass companion and for smaller $e_{\rm b}$ \citep[e.g.,][]{al94}.
  If the model by \citet{khi14} is adopted, the final semimajor axis of the planet 
  settles down to the location just inside the disk truncation radius, which is close to the stability limit.
  This result is not specific to the HD131399 system, 
  but can be applied generally to circumstellar gas giants in multiple stellar systems.  
  We use the HD131399 system as a reference system.  
  Note that circumstellar gas giant planets, $\gamma$ Cephei Ab, HD196885Ab, and HD41004Ab,
  in closer binary systems ($a_{\rm b} \sim 20$ au) also have orbits close to the stability limit
  \citep[e.g.,][]{Thebault_Haghighipour16}.
  Because these gas giants are at $\sim 2$ au, they can be in principle formed by
  the standard core accretion model.
  However, it is not clear if planet formation proceeds in such highly perturbed environments 
  \citep{Thebault_Haghighipour16}.
  These planets could also be formed by the model here,
  if another planet with $\ga$ 20-30 Earth masses existed at $\la 1$ au, as discussed later.
  
We briefly summarize the model of \citet{khi14} in Section 2.1 and
describe the assumptions and initial conditions for the discussions on 
HD131399 system in Section 2.2.
We present the results of numerical simulations in Section 3.
Section 4 is devoted to the summary and discussion.

\section{MODEL}
\subsection{The model by \citet{khi14}}

The scenario proposed by \citet{khi14} is based on the conventional 
core accretion model as follows:
(i) an icy core (or a core with some amount of gaseous envelope)
accretes from planetesimals in
inner regions at semimajor axis $\la$ 30 au, 
(ii) it is scattered outward by a nearby gas giant or more massive core to acquire a highly 
eccentric orbit with a periastron distance close to the original semimajor axis,
(iii) its orbit is circularized with significant lift-up of the periastron distance
through accretion of local protoplanetary disk gas along the
eccentric orbit, and (iv) through the local gas accretion, the planet becomes a gas giant.
For a given initial high eccentricity and semimajor axis,
\citet{khi14} investigated the details of the process in step (iii),
assuming that
the gas disk motion is Keplerian,  
the motions of the planet and the gas disk are coplanar
(they found that if the inclination is smaller than 30$^\circ$,
the final values of eccentricity and semimajor axis change by less than 5\%), 
the gas accretion rate onto the planet from local disk regions is independent of
instantaneous locations of the orbit, 
and the gas disk has a finite size (no gas accretion beyond the disk's outer edge).  

Here, we summarize their results (for detailed derivations of formulas, refer to \citet{khi14}).
The changes of the energy and angular momentum are derived 
analytically, although it must be solved numerically.
The damping rates of eccentricity $e$ and semimajor axis $a$ as a function of
the planetary mass $M$ are given by
\begin{equation}
  \frac{de}{d\log M}
  =-\frac{1-e^2}{e}\left(f_{\ell}(e, u_{\rm d})+\frac{1}{2}f_{\epsilon}(e, u_{\rm d})\right),
  \label{eq:dedm}
\end{equation}
\begin{equation}
  \frac{d\log a}{d\log M}=-f_{\epsilon}(e, u_{\rm d}),
  \label{eq:dadm}
\end{equation}
where $u_{\rm d}$ is the maximum eccentric anomaly ($0 <u_{\rm d} <\pi$)
within the gas disk of the outer edge at $r_{\rm d}$,
which is given by
\begin{eqnarray}
  u_{\rm d}\equiv
  \left\{
  \begin{array}{ll}
    \cos^{-1}\left[\frac{1}{e}\left(1-\frac{r_{\rm d}}{a}\right)\right]
    &\;\;\mbox{for}\;\;[Q > r_{\rm d}]\\
    \pi
    &\;\;\mbox{for}\;\;[Q < r_{\rm d}]
  \end{array}
  \right.
  ,
\end{eqnarray}
where $Q$ is the apoastron distance of the planet.
The functions $f_{\ell}$ and $f_\epsilon$ are
the change rates of
the specific angular momentum and orbital energy
over one orbital period
expressed as
\begin{equation}
  f_\ell = \int_{-t_{\rm d}/2}^{t_{\rm d}/2}s\left(
  \frac{l_{\rm gas}}{l_{\rm p}}-1
  \right)dt,
  \label{eq:fell}
\end{equation}      
\begin{equation}
  f_\epsilon = \int_{-t_{\rm d}/2}^{t_{\rm d}/2}s\left(
        \frac{\epsilon_{\rm gas}}{l_{\rm p}}-
        \frac{\epsilon_{\rm coll}}{l_{\rm p}}- 1\right)dt,
        \label{eq:fep}
\end{equation}
where
$t_{\rm d}$ is a duration at $r<r_{\rm d}$, $r$ is
the instantaneous distance of the planet from the central star,
$l_{\rm p}$ and $\epsilon_{\rm p}$ are the specific angular
momentum and orbital energy of the planet,
$l_{\rm gas}$ and $\epsilon_{\rm gas}$ are
those of accreting gas at instantaneous locations along the orbit,
and $\epsilon_{\rm coll}$ is energy dissipation by collision
between the planet and accreting gas.
The factor $s$ is given as follows.
The relative velocity between the planet and the unperturbed gas flow (circular Keplerian flow),
is readily derived by
\begin{eqnarray}
  v_{\rm rel}(r)^2 
  &=& \frac{GM_\ast}{r} \left( 3 - \frac{r}{a} - 2\sqrt{\frac{a}{r} (1-e^2) } \right),
\label{eq:vrel} 
\end{eqnarray}
where $M_\ast$ is the host star mass and $a$ and $e$ are
the semimajor axis and the eccentricity of the instantaneous planetary orbits.
However, because we are concerned with damping of initially highly eccentric orbits,  
the reduction of the relative velocity due to bow shock that occurs in front of the planet must be taken
into account.
\citet{khi14} derived the reduction factor $s$ for post-shock relative velocity ($s v_{\rm rel}$) 
with a 1D plane-parallel approximation as 
\begin{equation}
  s =\frac{(\gamma-1){\cal M}^2+2}{(\gamma+1){\cal M}^2}
  \simeq \frac{1}{4} \left( 1+ \frac{3}{{\cal M}^2} \right),
  \label{eq:s}
\end{equation}
where 
$\gamma$ is the specific heat ratio ($\gamma=7/5$ for H$_2$ molecules) 
and ${\cal M}$ is Mach number of pre-shock,
\begin{equation}
  {\cal M}=\frac{v_{\rm rel}}{c_s}
  = 30 \left( \frac{r}{1{\rm au}} \right)^{-1/4} \left( 3 - \frac{r}{a} - 2\sqrt{\frac{a}{r} (1-e^2) } \right)^{1/2},
\end{equation}
where we used an optically thin disk temperature, $T = 280(r/1{\rm au})^{-1/2}{\rm K}$, for evaluation of sound velocity $c_s$.

\subsection{Initial Conditions or Settings}

We apply Kikuchi et al.'s model to a reference planet, HD131399Ab.
We first evaluate the initial orbital elements and mass of a precursor body for HD131399Ab
and the gas disk that are substituted into Kikuchi et al.'s model,
where "initial" means the moment just after the scattering by an unseen planet.
We do not simulate the scattering process itself, but evaluate a reasonable range of the 
"initial" parameters after the scattering.

The initial periastron distance $q_{\rm i}$ must be around the unseen massive planet.
In the conventional core accretion model, $q_{\rm i}$ may be $\la 40$ au.
The upper limit of the initial apoastron distance $Q_{\rm i}$ may be limited by the existence of the binary companions.
The best$-$fit of the orbit of the binary by \citet{w16} is $a_{\rm b}=349\pm 28$ au and $e_{\rm b}=0.13\pm 0.05$ where $a_{\rm b}$ and $e_{\rm b}$ are the semimajor axis and eccentricity of the binary with respect to HD131399A.
We assume $a_{\rm b}=350$ au and $e_{\rm b}=0.13$ and consider the cases of $Q_{\rm i} \la 300$ au.
  The initial semimajor axis $a_{\rm i}$ and eccentricity $e_{\rm i}$ of
  the planet are calculated by $q_{\rm i}$ and $Q_{\rm i}$ as
\begin{eqnarray}
a_{\rm i} & = & \frac{Q_{\rm i}+ q_{\rm i}}{2}, \\
e_{\rm i} & = & \frac{Q_{\rm i}-q_{\rm i}}{Q_{\rm i}+q_{\rm i}}. \label{eq:ei}
\end{eqnarray}
The inclination to the gas disk is assumed to be zero for simplicity.
The initial mass of the planet ($M_{\rm i}$) must be larger than the critical core mass 
  for the onset of gas accretion.
  We use $M_{\rm i}=20 M_{\rm E}$ and 50 $M_{\rm E}$, where $M_{\rm E}$ is an Earth mass.
  The final mass of the planet is given by the observational best-fit value,
  $4 M_{\rm J}$ \citep{w16}.
  We stop the calculation when the mass of the planet reaches the final mass.
  In Kikuchi et al.'s model, orbital evolution is described by the planetary mass evolution but not by time evolution.
  We do not need to assume gas accretion rate onto the planet.
   
We use 100 au and 150 au for the gas disk radius 
-- because the circumstellar disk is tidally truncated by the perturbations 
from the binary companions --
 at $\sim 1/3$ of the distance between the primary star HD131399A and the binary
  \citep[e.g.,][]{al94}.
  The disk truncation radius is comparable to 
    or slightly larger than
    the long-term orbital stability limit radius.
We assume the gas disk exists until the end of the calculation.

\section{RESULTS}

\begin{figure}[hbtp]
  \begin{center}
    \resizebox{15cm}{!}{\includegraphics{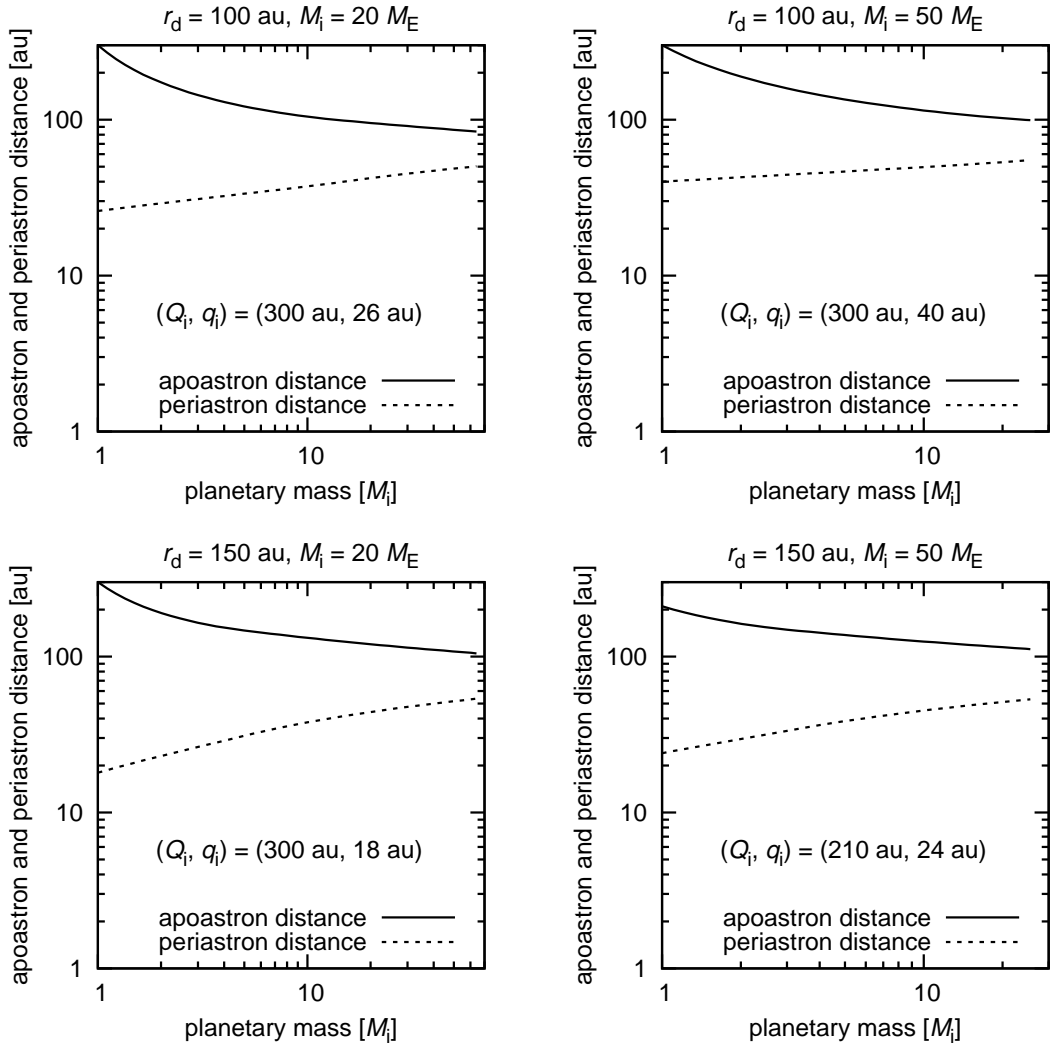}}
    \caption{
      Examples of evolution of the apoastron $Q$ and the periastron $q$ of a planetary orbit 
      as the growth of the planetary mass ($M$) for $r_{\rm d}$ = 100 au and 150,
      and $M_{\rm i} = 20 M_{\rm E}$ and $50 M_{\rm E}$.
      The initial conditions are $(Q_{\rm i} [{\rm au}], q_{\rm i}[{\rm au}])$
      = (300, 25), (300, 40), (300, 18), and (210, 24) from the top left to bottom right,
      respectively.
      The solid lines and dashed lines represent
      the apoastron and periastron distances, respectively.
    }
    \label{fig:ki_m_apq}
  \end{center}
\end{figure}

Using the initial semimajor axis and eccentricity,
we calculate the final semimajor axis and eccentricity for different values of
disk size $r_{\rm d}$ and the initial core mass $M_{\rm i}$,
integrating Eqs.~(\ref{eq:dedm}) and (\ref{eq:dadm}) 
  up to $M=4 M_{\rm J}$.
Figure \ref{fig:ki_m_apq} shows some examples of orbital evolution 
as a function of the planetary mass growth.
According to the mass growth of the planet, its eccentricity and semimajor axis are damped and
periastron $q$ and apoastron $Q$ converge.
Periastrons are lifted up quickly and the scattered planet becomes isolated from the perturber, the more massive planet.
  Apoastrons $Q$ settle down to $f r_{\rm d}$ with $f \sim 0.7-1.0$.
  \citet{khi14} showed that the final $Q$ is smaller for smaller $q_{\rm i}/r_{\rm d}$
  and smaller final $e$.
  Because the final $e$ is similar in each panel of Figure \ref{fig:ki_m_apq},
  the difference of the final values of $Q$ reflects the different values of
  $q_{\rm i}/r_{\rm d}$.
  
  The damping of $Q$ is associated with the planetary mass increase.
  This figure shows that the mass increase by 10 times from the initial value is
  enough to damp the apoastrons $Q$ to $\la r_{\rm d}$.
  The mass increase factor to the "final" state is $4 M_{\rm J}/M_{\rm i} = 25$ or 64.
  Thus, the planets' final orbits are inevitably close to the disk truncation radius.
  Orbital eccentricities are damped significantly from the initial values $\sim 1$,
  but have not been completely damped in this case.
  The semimajor axises are slightly smaller than $Q$ in the final state,
  corresponding to the retained $e$.
  Because the orbital stability limit is slightly closer to the primary star
  than
  the disk outer edge is, 
  the final semimajor axes are comparable to the orbital stability limit.
  With $Q \sim f  r_{\rm d}$,
  the final $a$ would be
  \begin{equation}
    a = (1-e)Q \sim (1-e) f r_{\rm d} \sim (1-e) (f/3)
    \left(\frac{r_{\rm d}}{a_{\rm b}/3}\right) a_{\rm b}.
    \label{eq:a_final}
  \end{equation}
  For $e\sim 0.35$, $f \sim 0.7-1.0$ and $r_{\rm d} \sim a_{\rm b}/3$,
  $a\sim (0.15-0.22) a_{\rm b}$.

\begin{figure}[hbtp]
  \begin{center}
    \resizebox{15cm}{!}{\includegraphics{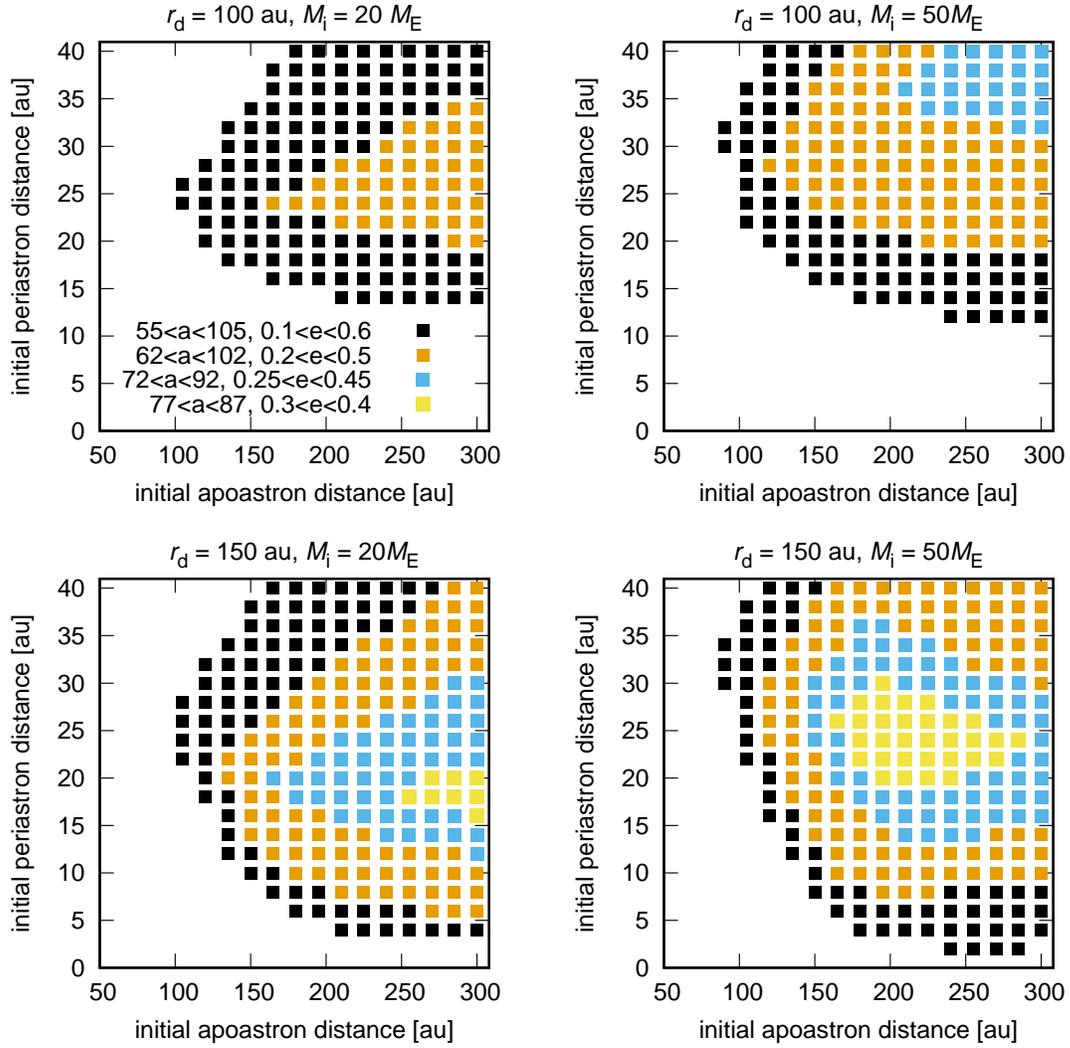}}
    \caption{
      Compatibility of the final semimajor axis $a$ and eccentricity $e$ of the planet 
      (when the planetary mass $M$ reaches 
      4 $M_{\rm J}$) to observationally estimated values
      of HD131399Ab as a function of initial $Q_{\rm i}$ and $q_{\rm i}$, 
      for $r_{\rm d}=100$ au and 150 au, and $M_{\rm i}=20 M_{\rm E}$ and $50 M_{\rm E}$.
      The yellow color points represent the best$-$fit: $77 {\rm au} < a < 87 {\rm au}$ and $0.3 < e < 0.4$.
      The other colors represent more deviated values. 
      Light blue: $72 {\rm au} < a < 92 {\rm au}$ and $0.25 < e < 0.45$,
      orange: $62 {\rm au} < a < 102 {\rm au}$ and $0.2 < e < 0.5$, 
      black: $55 {\rm au} < a < 105 {\rm au}$ and $0.2 < e < 0.5$, respectively.
    }
    \label{fig:ki_map}
  \end{center}
\end{figure}

Figure \ref{fig:ki_map} shows the values
of the semimajor axis $a$ and eccentricity $e$ of the planet 
when the planetary mass reaches 4 $M_{\rm J}$, 
on the $Q_{\rm i}-q_{\rm i}$ plane with available ranges of
50 ${\rm au} < Q_{\rm i} < 300$ au　and $q_{\rm i} < 40$ au for disk sizes,
$r_{\rm d}=100$ au and　150 au, and the initial planetary masses,
$M_{\rm i}=20 M_{\rm E}$ and $50 M_{\rm E}$.
  The parameter $f$ is controlled by $q_{\rm i}/r_{\rm d}$,
  so that $q_{\rm i}/r_{\rm d}$ and $r_{\rm d}/a_{\rm b}$ 
  (the binary separation is fixed to be $a_{\rm b}=350$ au) determine the final $a$,
  as shown in Eq.~(\ref{eq:a_final}).
  Because the mass increase factor determines the final $e$ and the final mass
  is fixed to $4 M_{\rm J}$, $M_{\rm i}$ determines the final $e$ in this case.
Each square on the map corresponds to one calculation that we have performed.
The observationally estimated semimajor axis and eccentricity of HD131399Ab are 
$82^{+23}_{-27}$ au and $0.35\pm 0.25$, respectively \citep{w16}.
Different colors for the squares represent how 
the calculated final values of $a$ and $e$ are consistent with 
the observational estimates, from yellow, light blue, orange, and black.
The yellow squares show the best ones, 
where $77 {\rm au} < a < 87 {\rm au}$ and $0.3 < e < 0.4$.
Figure \ref{fig:ki_map} shows that the current orbit of HD131399Ab is 
reproduced from the initial conditions with $q_{\rm i} \sim 20$-30 au,
a relatively large initial mass ($M_{\rm i}=50 M_{\rm E}$), and relatively
large disk truncation radius ($r_{\rm d}=150$ au).

If $Q_{\rm i}$ is close to the separation of the binary companion,
there is a risk that the planet is again scattered by the
binary companion pair until $Q$ is sufficiently damped.
For $M_{\rm i}=50 M_{\rm E}$ and $r_{\rm d}=150$ au,
the yellow squares exist down to $Q_{\rm i} \sim 150$ au, where the risk may be low. 
The relatively massive initial mass, $M_{\rm i}=50 M_{\rm E}$,  corresponds to
a planet on its way to becoming a gas giant.
In this case, a gas giant more massive than Saturn must have existed
at 20-30 au to scatter out the planet.

\section{SUMMARY AND DISCUSSION}

We have investigated the formation of circumstellar wide-orbit gas giants in multiple star systems,  
motivated by the 
announcement of the discovery of planet HD131399Ab by direct imaging \citep{w16}.
This planet is a circumstellar gas giant in a hierarchical triple star system and 
has perplexing orbital features.
The semimajor axis is about 1/4 of the separation between A and the B/C binary pair,
which means that the planet's orbit would be marginally stable or possibly unstable on long timescales. 
We have applied the formation model for wide-orbit gas giants by \citet{khi14} and found that 
the orbit close to the stability limit is not a fortunate accident, but rather an inevitable result.  

The suggested scenario is as follows:
(i) a core with or without a gaseous envelope is scattered by a nearby more massive planet
 at 20-30 au so that it develops a highly eccentric orbit with the periastron at 20-30 au, 
(ii) the scattered planet spends most of its time near the apoastron
on a highly eccentric orbit after the scattering, where the planetesimal accretion rate 
must be very small,
(iii) rapid gas accretion starts because of the low planetesimal accretion rate,
and the gas accretion is most efficient near the apoastron,
(iv) through gas accretion onto the planet, the orbit is circularized such that
the periastron distance is significantly increased and the apoastron distance
is adjusted to the tidal truncation radius of the disk by the binary companion ($r_{\rm d}$),
which is comparable to the orbital stability limit.
Therefore, this model is a promising option for 
to reproducing the current orbital configuration of HD131399Ab.
The best-fit results are obtained by the initial scattering location $q_{\rm i} \sim 20 $-30 au,
the initial planet mass $M_{\rm i}\sim 50 M_{\rm E}$ and the disk truncation radius $r_{\rm d} \sim 150$ au.

  If the orbital data of HD131399Ab is improved, the predicted initial conditions
  are better constrained.
  Even if HD131399Ab turns out be a background star, our model can be generally applied 
  to circumstellar gas giants in multiple stellar systems, 
  and it predicts that the planets would be often be located close to the stability
  limit against the perturbations of a companion(s).
  The results depend on the scaled parameters,
  $q_{\rm i}/r_{\rm d}$ and $r_{\rm d}/a_{\rm b}$, where $a_{\rm b}$ is the binary separation
  and $q_{\rm i}$ corresponds to the original semimajor axis of the planet
  before scattering.  
  The parameter $r_{\rm d}/a_{\rm b}$ is usually $\sim 1/3$, except for 
  the cases with high $e_{\rm b}$ or a very different mass companion from the host star,
  and the final $Q$ does not change sensitively for $q_{\rm i}/r_{\rm d}\la 0.1$ \citep{khi14}.   

  Other circumstellar gas giant planets close to the stability limit have already been
  discovered in more compact binary systems ($a_{\rm b} \sim 20$ au): 
  $\gamma$ Cephei Ab, HD196885Ab, and HD41004Ab, all of which
  are located at $\sim 2$ au.
  The parameter $r_{\rm d}/a_{\rm b}$ for these systems should be similar to HD131399Ab.
  Our model can reproduce their orbits close to the stability limit 
  with reasonable ranges of values of $q_{\rm i}$ and $M_{\rm i}$,
  while these planets could also be formed by the standard in situ core accretion model.
  Note that our model needs another planet with $\ga 20-30M_{\rm E}$ that scattered
  the core, which has not been detected in the $\gamma$ Cephei A, HD196885A,
  and HD41004A systems.
  Furthermore, in compact binary systems,
  there is a greater risk that
  the scattered core will
  be perturbed by the binary companion before the scattered orbit shrinks.
  For very wide binary systems, our model may not work well either.
  In our model, $Q_{\rm i}$ must be between $r_{\rm d} \sim a_{\rm b}/3$ and $a_{\rm b}$,
  while it is likely that $q_{\rm i} \la 30$ au.
  From Eq.~(\ref{eq:ei}),  $1 > e_{\rm i} \ga (a_{\rm b} - 90)/(a_{\rm b} + 90)$.
  The range of $e_{\rm i}$ that satisfies this condition becomes narrower as
  $a_{\rm b}$ increases.
  Furthermore, the disk size cannot become infinitely large when $a_{\rm b}$ is
  very large.
  In this case, $r_{\rm d}$ is independent of the stability limit.
  Thus, our model is better applied to binary systems with separations of,
  say, 100-1000 au, while it could also be applied to more compact or more
  extended binary systems.

  It is also pointed out that {\it circumbinary} gas giant planets tend to be located near
  the {\it inner limit} of the stable region around the binary systems
  because the curcumbinary disks are truncated 
  by an inner binary pair and planetary migrations stop at the inner edge \citep{kh14,kh15}.
  Here, we predict that {\it circumplanetary} wide-orbit gas giant planets 
  tend to be located near the {\it outer limit} of the stable region around the host star.

These predictions should be tested by more samples of direct imaging observations 
for wide-orbit gas giants in multiple stellar systems.

\acknowledgements
  This paper was motivated by a group project at Lorentz center workshop
  "New Directions in Planet Formation," held in 2016 July, that S.I. participated in.
  S.I. thanks the team members of the group project, Nader Haghighipour, Judit Szul\'agyi,
  Steven Rieder, Katherine Kretke, and Jiwei Xie.
We thank Akihiro Kikuchi for providing us the code developed
for our previous paper, \citet{khi14} 
  and an anonymous referee for useful comments.
Data analyses were in part carried out on the PC cluster at 
the Center for Computational Astrophysics, 
National Astronomical Observatory of Japan.
This work was supported by JSPS KAKENHI grant 15H02065.

\begin{thebibliography}{} 


\bibitem[Artymowicz \& Lubow (1994)]{al94}
  Artymowicz, P. \& Lubow, S. H. (1994)
  \apj 421, 2, 651

\bibitem[Bromley \& Kenyon (2014)]{bk14}
  Bromley, B. C. \&  Kenyon, S. J. (2014)
  \apj 796, 2, id.141

\bibitem[Ida \& Lin (2004)]{il04}
  Ida, S. \& Lin, D. N. C. (2004)
  \apj, 604, 388

\bibitem[Holman \& Wiegert(1999)]{holman99}
 Holman, P. \& Wiegert, W. (1999)
  \aj, 117, 621  

\bibitem[Kalas et al. (2008)]{k08}
  Kalas, P.,  Graham, J. R., Chiang, E., Fitzgerald, M. P., 
  Clampin, M., Kite, E. S., Stapelfeldt, K.,
  Marois, C., \& Krist, J. (2008)
  Science, 322, 5906

  
\bibitem[Kozai (1962)]{k62}
  Kozai, Y. (1962)
  \aj, 67, 591
  

\bibitem[Kikuchi et al.~(2014)]{khi14}
  Kikuchi, A., Higuchi, A., \& Ida, S. (2014)
  \apj, 797, 1, id.1

  
\bibitem[Kley \& Haghighipour(2014)]{kh14}
  Kley, W., \& Haghighipour, N. (2014)
  \aap, 564, id.A72


\bibitem[Kley \& Haghighipour(2015)]{kh15}
  Kley, W., \& Haghighipour, N. (2015)
  \aap, 581, id.A20

\bibitem[Kuzuhara et al.(2013)]{k13}
  Kuzuhara, J., Thalmann, C., Janson, M., 
  Kandori, R.,  Brandt, T. D.,  Thalmann, C.,  Spiegel, D.,  Biller, B.,  
  Carson, J.,  Hori, Y.,  Suzuki, R.,  Burrows, A.,  Henning, T.,  
  Turner, E. L.,  McElwain, M. W.,  Moro-Martín, A.,  Suenaga, T.,  
  Takahashi, Y. H.,  Kwon, J.,  Lucas, P.,  Abe, L.,  Brandner, W.,  Egner, S.,  
  Feldt, M.,  Fujiwara, H.,  Goto, M.,  Grady, C. A.,  Guyon, O.,  Hashimoto, J.,  
  Hayano, Y.,  Hayashi, M.,  Hayashi, S. S.,  Hodapp, K. W.,  Ishii, M.,  Iye, M.,  
  Knapp, G. R.,  Matsuo, T.,  Mayama, S.,  Miyama, S.,  Morino, J.-I.,  
  Nishikawa, J.,  Nishimura, T.,  Kotani, T.,  Kusakabe, N.,  Pyo, T.-S.,  
  Serabyn, E.,  Suto, H.,  Takami, M.,  Takato, N.,  Terada, H.,  Tomono, D.,  
  Watanabe, M.,  Wisniewski, J. P.,  Yamada, T.,  Takami, H., \& Usuda, T.
  \apjl, 763, L32



\bibitem[Lidov(1962)]{l62}
Lidov, M. L. (1962)
Planetary and Space Science, 9, 10

\bibitem[Marois et al.(2008)]{m08}  
  Marois, C.,  Macintosh, B.,  Barman, T.,  Zuckerman, B.,  Song, I.,  Patience, J.,  
  Lafreni$\acute{\rm e}$re, D., \& Doyon, R. (2008)
  Science, 322, 5906
  
  
\bibitem[Thebault \& Haghighipour(2016)]{Thebault_Haghighipour16}   
Thebault, Ph. \& Haghighipour, N. (2016)
in Planetary Exploration and Science: Recent Advances and Applications, eds. S. Jin, N. Haghighipour, W.-H. Ip, Springer

  
\bibitem[Veras et al.(2017)]{veras17}
  Veras, D., Mustill, A. \& Gänsicke, B. T.
 (2017) MNRAS 465, 1499


\bibitem[Wagner et al.(2016)]{w16}
  Wagner, K., Apai, D., Markus Kasper, M.,
  Kratter, K., McClure, M., Robberto, M., 
  \& Beuzit, J. (2016)
  Science, 353, 6300

\end{thebibliography}
\end{document}